\documentclass{ws-p8-50x6-00}

\begin{document}

\title{The COMPASS Experiment and the measurement of the gluon
polarisation }

\author{F. Tessarotto}

\address{CERN (CH) and INFN-Trieste (I)\\ 
E-mail: fulvio.tessarotto@cern.ch}

\maketitle

\abstracts{ COMPASS,
a new fixed target experiment at CERN, aimed at the study of nucleon
spin structure and hadron spectroscopy, has started to collect physics
data in~Autumn 2001.
This paper describes the COMPASS apparatus and the measurement of the
gluon polarisation $\Delta G /G$. \\
The apparatus consists in a solid state polarised
target and a two stage spectro-meter with high resolution tracking,
particle identification and calorimetry, capable of standing high
event rates. \\
COMPASS measures the longitudinal spin asymmetry of open charm production
in polarised deep inelastic muon nucleon scattering: this asymmetry is
directly related to $\Delta G$ since at COMPASS energies open charm is
essentially produced by photon gluon fusion only. A second channel,
used to access $\Delta G$ with higher statistics,
is the production of correlated high $p_T$ hadron pairs.}

\section{Introduction}
A large experimental effort
at CERN, SLAC and DESY during the last decade
led to the collection of high quality inclusive data on polarised
deep inelastic scattering (DIS) and several NLO QCD analyses have
extracted accurate quark helicity distributions. Despite of many attempts
(including a comparison between high $p_T$ data and monte carlo simulations
by HERMES\cite{hermes})
the gluon helicity distribution $\Delta G$ is still substantially
unconstrained but most indications suggest positive large values.
\par
In the incoming years the effort to clarify
the nucleon spin puzzle will benefit from direct measurements of $\Delta G$
by a new generation of experiments:
COMPASS (NA48) at CERN SPS, STAR and PHENIX at BNL
Polarised Proton Collider and 
E161 (Real Photon Experiment) at SLAC.

\section{The COMPASS Collaboration}
\par
In 1996 the two communities which had presented the HMC and CHEOPS
Letters of Intent for fixed target experiments at CERN merged in the
COMPASS (COmmon Muon and Proton Apparatus for Structure and Spectroscopy)
Collaboration and presented a Proposal \cite{prop} which obtained
approval in 1997.
35 Institutes participate in the Collaboration,
for a total of almost 200 physicists. \\
COMPASS has a broad physics program with different beams and targets.
It has a spin program with polarised muon beam: its main goal is to provide
the first direct measurement of the gluon polarisation $\Delta G /G$;
it will perform flavour decomposition of
the quark helicity distributions; determine the
transversity structure function $h_1$; measure polarised fragmentation
functions.
%like $\Delta {\cal D}_q^{\Lambda}$.
\par
Using hadronic beams it will
study Primakoff reactions to obtain $\pi $ and $K$ polarisabilities;
perform extensive meson spectroscopy to investigate
the presence of exotic states (glueballs or hybrids);
collect large samples of semi-leptonic decays of charmed mesons and baryons
to measure the CKM matrix elements $V_{cs}$ and $V_{cd}$, determine 
formfactors and probe predictions from Heavy Quark Effective Theory; 
perform a systematic study of charm hadroproduction cross sections;
observe doubly charmed baryons for the first time.
\section{The COMPASS Apparatus}
For the spin program COMPASS uses the $ \mu^+ $ beam from the CERN SPS,
with an
energy between 100 and 200 GeV. Muons are produced
by parity violating decay of $ \pi $ (and K) mesons and are
naturally polarised: $ P_B \approx -80 \% $. The beam intensity is
$ 2.2 \cdot 10^8 ~\mu^+ $ per spill, with 5 s long spills of 14.4 s period.
Each incoming muon is tracked by scintillating fibres
hodoscopes and momentum analysed before the target.

\par
The target consists in two cells (60 cm long, 3 cm diam.)
filled with solid state
NH$_3$ for proton and $^6$LiD for deuteron measurements.
The luminosity $ {\cal L} \approx 5 \cdot 10^{32} ~cm^{-2}s^{-1}$
(for $^6$LiD).
%\par
A $^3$He -$^4$He dilution refrigerator keeps the target at T $<$
100 m$^{\circ}$K, while a solenoid provides a 2.5 T magnetic
field along the beam axes; a dipole is used for the rotation of
the polarisation.
\par
The COMPASS solenoid (600 mm diam.) is not yet available and this year
the solenoid from the SMC experiment (255 mm diam.) was used.
%resulting in a reduced angular acceptance.
\par
The two target cells are dynamically polarised
%(by microwave irradiation at
%frequencies close to the Larmor frequency of the paramagnetic centres
%present in the material)
and kept in opposite directions,
(either longitudinal or transverse);
the polarisation of the nucleons is measured via 10 NMR coils
with an accuracy of about 3\%.
In NH$_3$ protons can be polarised at $P_T \approx $ 90\%, and the
dilution factor is $f \approx $ 0.17, while for $^6$LiD, successfully
used in the 2001 run, $P_T \approx $50\% and $f \approx $ 0.5.
%since the $^6$Li nucleus can be considered as $^4He$ + D.
%\par
%For this run the COMPASS solenoid (600 mm diam.) was not available and
%the solenoid from SMC experiment (255 mm diam.) has been used,
%resulting in a reduced angular acceptance.

%\begin{figure}[b] %{l}  %{0.5\textwidth}
%\epsfig{figure=setup1.ps,width=8.5cm,height=6.5cm}
%\caption{\small The COMPASS setup for the run of year 2001.
%\label{setup}
%}
%\medskip
%\end{figure}

%\subsection{The Spectrometer}\label{subsec:spectrometer}
\par
%A common requirement of all foreseen measurements is the high resolution
%detection and identification of particles over a large angular and
%dynamical range.
To achieve high resolution detection and identification of particles
over a large angular and dynamical range the COMPASS
spectrometer comprises two magnetic stages: the first has an acceptance of
$ \pm $180 mrad and a large aperture dipole magnet providing 1 Tm
bending power; the second, with 40 mrad acceptance and 5 Tm bending power
is used to analyse high momentum particles.
\par
Tracking in the beam region is granted by a set of stations of
scintillating fibres hodoscopes, with fibres diameters ranging
from 0.5 mm in the central region to 1 mm in the outer regions, read by
multi-anode PMs: they provide 400 ps time resolution.
\par
The high flux area around the beam region is covered by micro-pattern
detectors: Micromegas and GEMs. COMPASS Micromegas \cite{MM}
contain a thin micro-mesh foil
%with 60\% optical transparency
separating an ionization
volume from a high  field (40 KV/cm, 100 $\mu m$ gap) amplification region;
they have high efficiency and a space resolution of $ \approx  80 \mu m$.
COMPASS GEMs (Gas Electron Multiplier) \cite{GEM} are made of
kapton foils having Cu layers on both sides and a large
($10^4 /cm^2$) number of $ 60 \mu m$ diameter holes: a high field inside
the holes, provides electrons amplification.
Triple GEM chambers are used with 31$ \times 31 cm^2 $ active area and
two-dimensional projective readout
with 100 $\mu m$ resolution.
%have been successfully used in the run of year 2001.
\par  
A set of MWPCs with fast electronic readout, planar drift chambers
and drift chambers
made of straw tubes layers with a size of 320$ \times 240 cm^2$ are
used for the large area tracking.
\par
Both stages of the spectrometer are equipped with high resolution
electro-magnetic and hadron calorimeters, muon filters
and hadron identification provided by a RICH detector. The high momentum
RICH is not yet available, while
the large acceptance RICH, RICH1 is fully equipped,
and consists in a 3.3 m long vessel filled with a heavy fluorocarbon
radiator gas ($ C_4 F_{10}$), a large (5. m high, 6. m wide)
wall of spherical mirrors which focuses
Cherenkov photons onto a set of UV photon detectors (MWPC's with quartz
windows and CsI photo-converting layers deposited on pcb cathodes
segmented in $8\times 8 mm^2$ pads) with signal
amplitude readout for a total of 83000 channels.
RICH1 should provide $\pi - K$ separation in the range
between 3 GeV/$c$ and 65 GeV/$c$.
\par
Coincidences between elements of hodoscope planes at different
positions along the beam select scattered muons for triggering purpose.
\par
Parallel readout front-end electronics with local pre event building
and pipelined acquisition system allow to stand trigger rates up to 100
kHz with minimal dead-time with event sizes $\approx $ 30 kB.
\par
COMPASS uses fully object oriented databases and software for storage
and analysis of the 35 MB/s data flow (300 TB/year stored data) and
has set-up a computer farm of $\approx $ 200 PCs.

\section{Measurement of $\Delta G$}
The gluon polarisation $\Delta G / G$ will be accessed by measuring
the asymmetry of two processes:
the open charm production and the correlated high $p_T$ hadron
pair production.
\vspace{-.5cm}
%\subsection{Open Charm production}\label{subsec:puzzle}
%\par
%\begin{tabular}{c c}
\begin{figure}[h]
\vspace{.2cm}
\begin{minipage}[t]{12pc}
%\figurebox{20pc}{15pc}
\epsfxsize=12pc 
\epsfbox{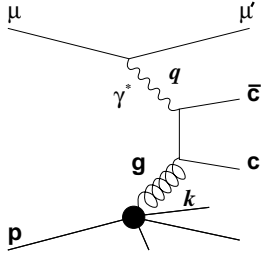}
\vspace{-.8cm}
\caption{The leading order diagram for open charm production:
the photon gluon fusion (PGF) process.
\label{fig:pgf}}
\end{minipage}
\end{figure}
%&
%\vspace{2cm}
%\hspace{6cm}
\begin{figure}[h]
\vspace{-6.7cm}
\hspace{6.6cm}
\begin{minipage}[t]{12pc}
\epsfig{figure=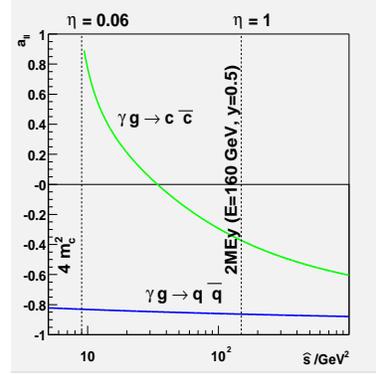,width=5.cm,height=5.cm}
\vspace{-.6cm}
\caption{The asymmetry of 
the PGF process in the photoproduction limit
\label{jpretz_asy_pgf}}
\end{minipage}
\end{figure}

Charm production in DIS is dominated by the 
photon gluon fusion process $ \gamma^{\ast} q \rightarrow c \bar c$.
%in which a virtual photon with four-momentum $q$ interacts with a gluon
%carrying a fraction $\eta$ of the parent nucleon momentum $p$.
The scale of the process is the photon-gluon centre of mass energy $\hat s$
%\begin{equation}
%\hat s = (q + \eta p)^2 \approx -Q^2 +2 \eta M E y
%\label{eq:s}
%\end{equation}
%where $Ey$ is the virtual photon energy and M is the nucleon mass.
which is large ($\hat s > 4 m_c^2$) even for
very low $Q^2$, allowing to make use of 
the whole production cross section,
%(up to 0.4\% of $\sigma_{tot}^{\gamma p}$);
which is both measured and
calculable: for 100 GeV beam and
energy transfers between 35 GeV and 85 GeV the COMPASS useful
cross sections is 1.9 nb.
The systematics from intrinsic charm and diffractive or resolved photon
contributions are small.
\par
The detection of open charm hadrons in the final state of an event will
be primarily done by identifying $D^0$ and $\overline{D^0}$ mesons through
their golden decay channel: $D^0 \rightarrow K^- \pi^+, \overline{D^0}
\rightarrow K^+ \pi^- $. On average 1.2 neutral $D$s are produced per
open charm event; the branching ration of this decay is 4\%.
\par
In a mass window of $m(D^0) \pm 20$ MeV the typical signal to background
ratio expected for COMPASS
is 1:30. By imposing a cut on the $K$ direction in the $D^0$ rest
frame: $ |cos (\theta^*_K)| < 0.5$ and on $z_D = E_D / E_{\gamma^*} < 0.25$
the ratio becomes 1:4, with an efficiency of $\approx $30\%.
This corresponds to almost 1000 reconstructed $D^0$/day.
\par
Further improvements will come from other $D^0$ decay
channels and other open charm channels, like $D^+ \rightarrow K^+ \pi^+ \pi^-$
(B.R. = 9\%) and in particular from the $D^{* \pm}\rightarrow D^0
\pi^{\pm}$ decay which is almost background free due to the small
mass difference: $m(D^*) - m(D^0)$ = 145 MeV. \\
The measured asymmetry is:
%\begin{center}
%{$A^{exp} =
{$({N^{\uparrow \downarrow}_{c \bar c}
 - N^{\uparrow \uparrow}_{c \bar c}}) / 
({N^{\uparrow \downarrow}_{c \bar c}
 + N^{\uparrow \uparrow}_{c \bar c}})
 = P_B P_T f D A^{\gamma^*N}_{c \bar c} $}
%\approx 0.1~A^{\gamma^{\star}N}_{c \bar c} $}
%\end{center}
where $N^{\uparrow \uparrow}_{c \bar c} (N^{\uparrow \downarrow}_{c \bar c})$
is the number of events with target spin
parallel (anti-parallel) to the $\mu$ spin and $D$ is the $\gamma^*$
depolarisation ($D \approx 0.66$).
\par
$\Delta G( \eta )$ can be extracted from $ A^{\gamma^*N}_{c \bar c} $
at an average value of the nucleon momentum fraction carried by the gluon
$\eta$=0.1 using:
\begin{center}
$\int_{4m_c^2}^{2MEy} \Delta \sigma(s)^{\gamma g \rightarrow c \bar c} 
      {\bf \Delta G(\eta,s)} { ds} = A^{\gamma^{\star}N}_{c \bar c} \cdot
\int_{4m_c^2}^{2MEy} 
      \sigma(s)^{\gamma g \rightarrow c \bar c} {G(\eta,s)} ds $
\end{center}
where the unpolarised terms can be taken from literature and
$\Delta \sigma(s)^{\gamma g \rightarrow c \bar c}$ comes from QCD calculations.
The partonic asymmetry $\Delta \sigma(\hat s) / \sigma(\hat s)$ is shown
in fig.~\ref{jpretz_asy_pgf} in the photoproduction limit ($Q^2=0$).
\par
For one year of running COMPASS expects from the open charm asymmetry
measurement a statistical error 
$ \sigma (\Delta G /G) \approx 0.15$.
\par
Fig.~\ref{jpretz_asy_pgf} shows that the partonic asymmetry for light quark
production in photon-gluon fusion is large too. In order to discriminate
this process from the leading order $\gamma^* q \rightarrow q$ process
events with hadron pairs having correlated high transverse
momentum  will be selected: ($p_T > 1$ GeV/$c$ and $\Phi_{h_1} - \Phi_{h_2}
= 180^o \pm 30^o $ where $\Phi_h$ is angle between the lepton
scattering plane
and the plane containing the virtual photon and the hadron momenta).
\par
A more abundant production rate for these events is expected in comparison
with the open charm events, and their kinematics allows the reconstruction of
$\eta$, probing $\Delta G(\eta)$ in different $\eta$ bins
in the range 0.02 $< \eta <$ 0.4.

\begin{figure}[t] %{l}  %{0.5\textwidth}
\hspace{2cm}
%\begin{minipage}[t]{12pt}
\epsfig{figure=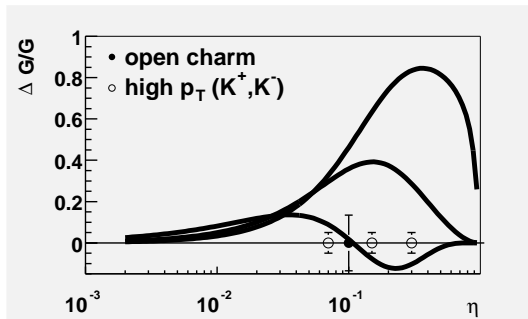,width=7.cm}
\caption{\small The statistical error on $ \Delta G(\eta) / G(\eta) $
for both the open charm method (full point) and the high $p_T$ hadron
pair method (open points) for one year.
%The three curves show parameterisation from Gehrmann and Stirling
%\cite{gs}.
\label{jpretz_dG}
}
\medskip
%\end{minipage}
\end{figure}

\par
A concurrent process which contributes for about 26\% of the events is the
QCD compton scattering $\gamma^* q \rightarrow qg$.
The selection of $K^+ K^-$ pairs will reduce its contribution and
the comparison of equal and opposite sign pairs can help to estimate it,
but the systematic errors will remain larger than in the open charm case.
\par
The expected statistical errors for one year of run for the two methodes
are presented in fig.~\ref{jpretz_dG} together with three parametrisations
of $\Delta G/G$ from Gehrmann and Stirling \cite{gs}: a measurement with
similar resolution will represent a major step toward a deeper
understanding of the nucleon spin.
\section{Conclusion}
COMPASS is a new fixed target experiment at CERN with a wide physics
program on hadron structure and spectroscopy: it has an outstanding
apparatus with high resolution tracking, particle identification and
calorimetry, capable of standing high event rates.
It will contribute to the large experimental
effort to clarify the spin structure of the nucleon by providing
the first
direct measurement of the gluon polarisation $\Delta G/G$ with an
accuracy of 10\%.


\begin{thebibliography}{99}

\bibitem{hermes}A. Airapetian {\it et al}, \Journal{\PRL}{84}{2584}{2000}.
\bibitem{prop} The COMPASS Collaboration, COMPASS Proposal, \\
{\em CERN/SPSLC 96-14} (March 1, 1996).
\bibitem{MM}D. Thers {\it et al}, \Journal{\NIM}{A469}{133}{2001}.
\bibitem{GEM}F. Sauli, \Journal{\NIM}{A386}{531}{1997}.
\bibitem{gs}T. Gehrmann and W.J. Stirling, \Journal{\PRD}{53}{6100}{1996}.

\end{thebibliography}
\end{document}